\documentclass[12pt]{iop/iopart}

\usepackage{iop/iopams}  
\usepackage{hyperref}
\usepackage{graphicx}
\usepackage{caption}
\DeclareUnicodeCharacter{2009}{\,}

\usepackage[backend=bibtex, maxnames=1,sorting=none]{biblatex}
\begin{document}

\title[]{Computing the Wave: Where the Gravitational Wave Community benefits from High-Energy Physics, and where it differs ?}

\author{Marco Meyer-Conde$^{1,2,3}$, Nobuyuki Kanda$^{1,4}$, Hirotaka Takahashi$^{2}$, Ken-ichi Oohara$^{5,6}$, Kazuki Sakai$^{7}$}
\vspace{10pt}
\begin{indented}
\item[]\textit{On behalf of the LIGO Scientific Collaboration, Virgo Collaboration and KAGRA Collaboration} (March 2024)
\end{indented}

\vspace{10pt}

\address{$^1$Department of Physics, Osaka Metropolitan University, Japan}
\address{$^2$Research Center for Space Science, Advanced Research Laboratories, Tokyo City University, Japan}
\address{$^3$University of Illinois at Urbana-Champaign, Department of Physics, USA}
\address{$^4$Nambu Yoichiro Institute of Theoretical and Experimental Physics (NITEP), Osaka Metropolitan University, Japan}
\address{$^5$ Graduate School of Science and Technology, Niigata University, Japan}
\address{$^6$ Niigata Study Center, The Open University of Japan}
\address{$^7$ Department of Electronic Control Engineering, NIT, Nagaoka College, Japan}
\vspace{10pt}
\ead{marco@tcu.ac.jp}

\begin{abstract}
High-Energy Physics (HEP) and Gravitational Wave (GW) communities serve different scientific purposes. However, their methodologies might potentially offer mutual enrichment through common software developments. A suite of libraries is currently being prototyped and made available at \url{https://git.ligo.org/kagra/libraries-addons/root}, extending at no cost the CERN ROOT data analysis framework toward advanced signal processing. We will also present a performance benchmark comparing the FFTW and KFR library performances.
\end{abstract}

%

\section{Introduction}

The timely detection of binary neutron star mergers holds paramount significance in both astrophysics and nuclear physics. The first of such kind in 2017, GW170817+GRB170817A~\cite{gw170817}, marked the beginning of multi-messenger astronomy and paved the way for studying neutron stars, predicted to have a sizable quark-matter core through both gravitational wave and its electromagnetic counterpart. For the last few years, the main GW collaborations joined their computing efforts and built a common infrastructure, serving the International Gravitational-Wave Observatory Network (IGWN)~\cite{igwn} led by LIGO~\cite{ligo}, Virgo~\cite{virgo}, and KAGRA~\cite{kagra} collaborations. Such infrastructure aims to extend and consolidate its computing infrastructure using novel technologies such as high-performance computing and cloud technologies. This transformative period is driven by the synergy of thousands of researchers and engineers to make new sciences. IGWN infrastructure and CERN Data Centre face significant common computing challenges, including integrating modern machine learning techniques, though their requirements differ. CERN notoriously requires substantial computing power to process the massive amounts of collected data. At the same time, IGWN focuses also on real-time data streams and low-latency measurements, aiming to support follow-up observations by astronomers and the analysis of transient astrophysical events.\\

This proceeding explores the common aspects of high-energy physics and the gravitational wave community, highlighting the intersections in their technological and computational challenges. It will focus on the advanced signal processing library, with a special highlight in comparing FFTW and KFR performances in a benchmark test.

\section{IGWN Computing Infrastructure}
\subsection{Low Latency Alert Generation Infrastructure}

\begin{figure}[h]
    \centering
    \includegraphics[width=0.8\textwidth]{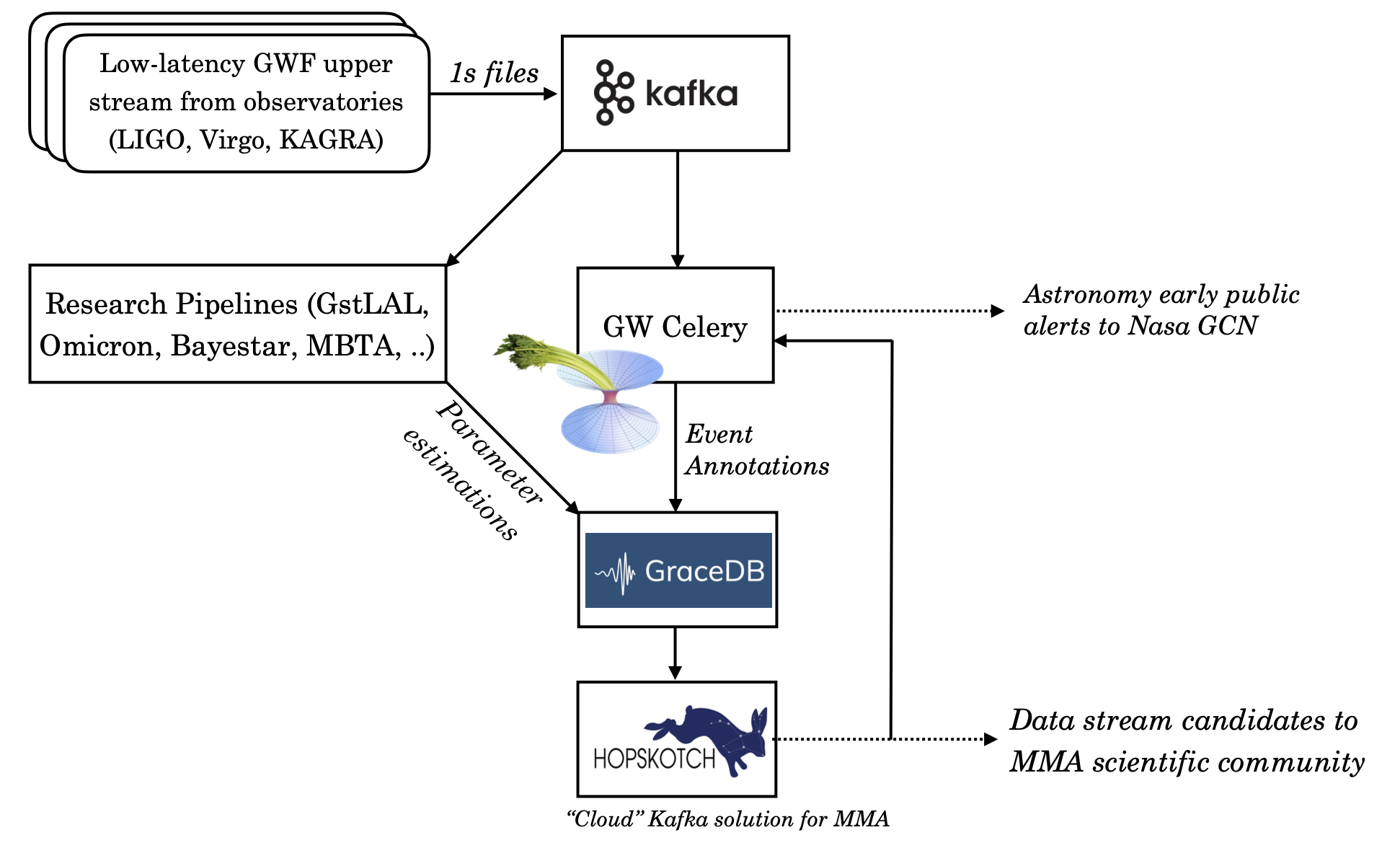}
    \caption{Low Latency Alert Generation Infrastructure Workflow Overview.}
    \label{fig:llai-workflow}
\end{figure}

The Low Latency Alert Generation Infrastructure (LLAI)~\cite{llai} is responsible for rapidly processing and issuing gravitational wave alerts following the workflow illustrated in Fig.~\ref{fig:llai-workflow}. It is optimized for real-time performance, ensuring significant events are communicated with minimal delay. A core element of this infrastructure is the GWCelery software~\cite{gwcelery}, which orchestrates the entire workflow from data ingestion to alert generation by tagging event candidates over time.

\begin{figure}[h]
    \centering
    \includegraphics[width=\textwidth]{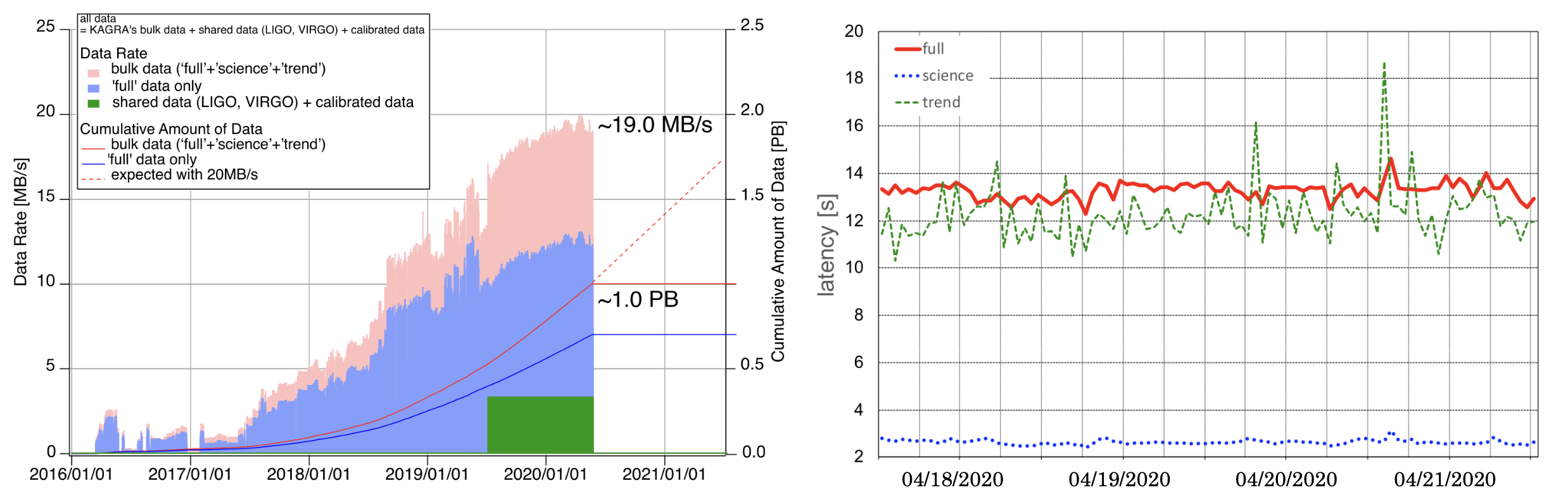}
    \caption{(a) Data volume projections at the time of O3 run; (b) latency estimate from Kamioka mines to Kashiwa ICRR cluster nearby Tokyo around the same time~\cite{kagra_dmg_ptep}.}
    \label{fig:kagra-volumes}
\end{figure}

LLAI utilizes a combination of dedicated clusters and cloud-based infrastructures. The CIT cluster in California offers substantial computational power, while cloud solutions like AWS clusters provide flexibility and scalability. At this time, the INFN-CNAF facility is also investigating the Kubernetes cluster to extend the capabilities of LLAI. Such hybrid approaches allow LLAI to adapt effectively to varying data volumes and computational needs. Fig.~\ref{fig:kagra-volumes} highlights the datastream rate and total latency going from the experimental side to the KAGRA main cluster in various dataset cases. Science case, as illustrated in Fig.~\ref{fig:kagra-volumes}b, is transferred with high priority to IGWN clusters.

\begin{figure}[h]
    \centering
    \includegraphics[width=0.9\textwidth]{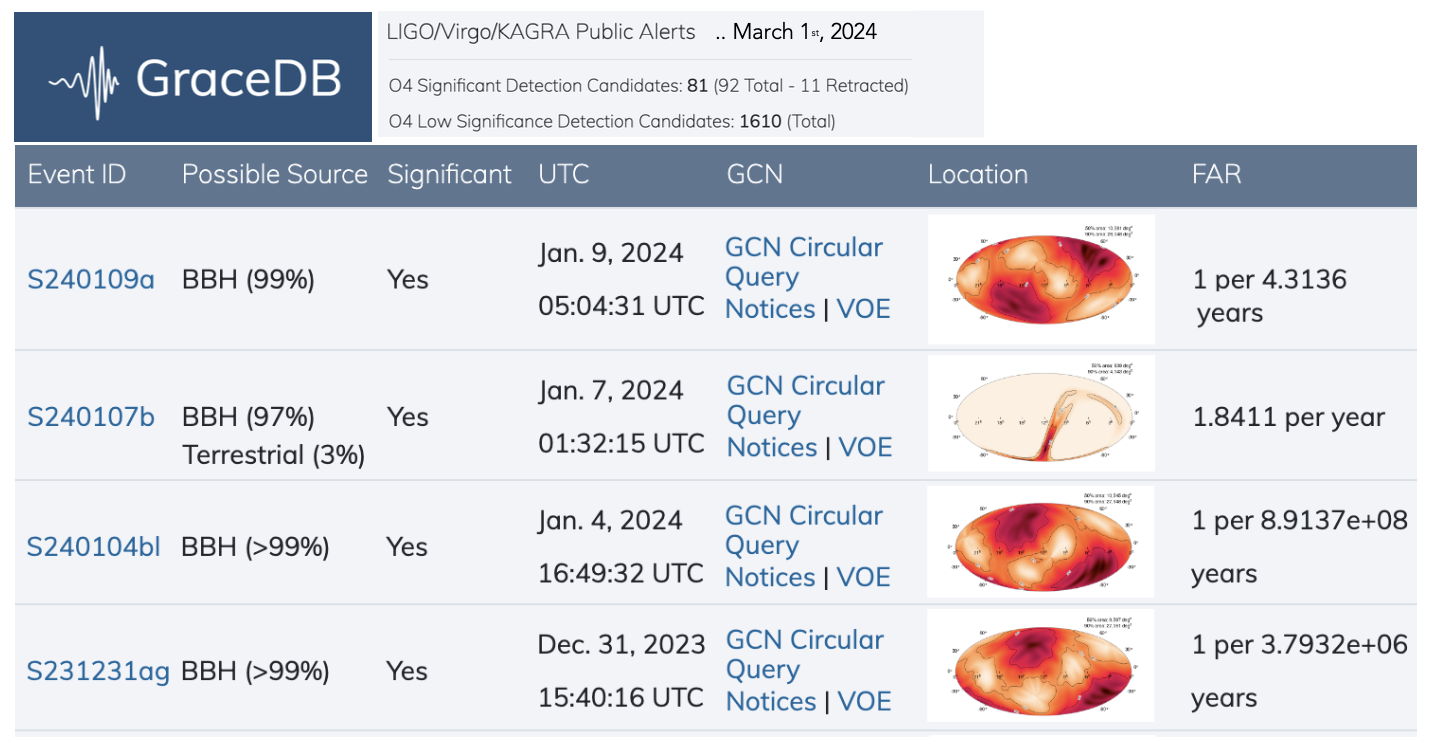}
    \caption{GraceDB Centralized Database Overview highlighting 81 significant detection candidates in O4 as compared to the 90 events in total through O1 to O3}
    \label{fig:gracedb}
\end{figure}

Apache Kafka~\cite{kafka} software is a critical component managing data streaming and message brokering at low latency. Gravitational wave data is continuously streamed into the system, providing a scalable and reliable method for handling high-speed throughput data leveraged by \textit{igwn-lldd-common} library~\cite{igwn-lldd-common}. Typical latencies between Kashiwa and KAGRA, LIGO, Virgo are in average 2.5~s, 6-8~s, 10~s, respectively. Moreover, GraceDB~\cite{gracedb} serves as the centralized database for event candidates and is accessible via a web service. It manages event data efficiently, marking relevant portions of the data stream by gathering the output of multiple research pipelines that are continuously running. Significant events are then prominently displayed and prioritized for immediate attention, as shown in Fig.~\ref{fig:gracedb}.

Eventually, event alerts are sent to NASA's Global Coordinates Network (GCN) and accompanied by circulars that provide updates and additional details~\cite{gcn}. This network ensures alerts reach the broader astronomical community. Early warnings and other astronomical observations are coordinated through SCiMMA Hopskotch~\cite{scimma}.

\subsection{Offline Analysis Infrastructure}

\begin{figure}[h]
    \centering
    \includegraphics[width=\textwidth]{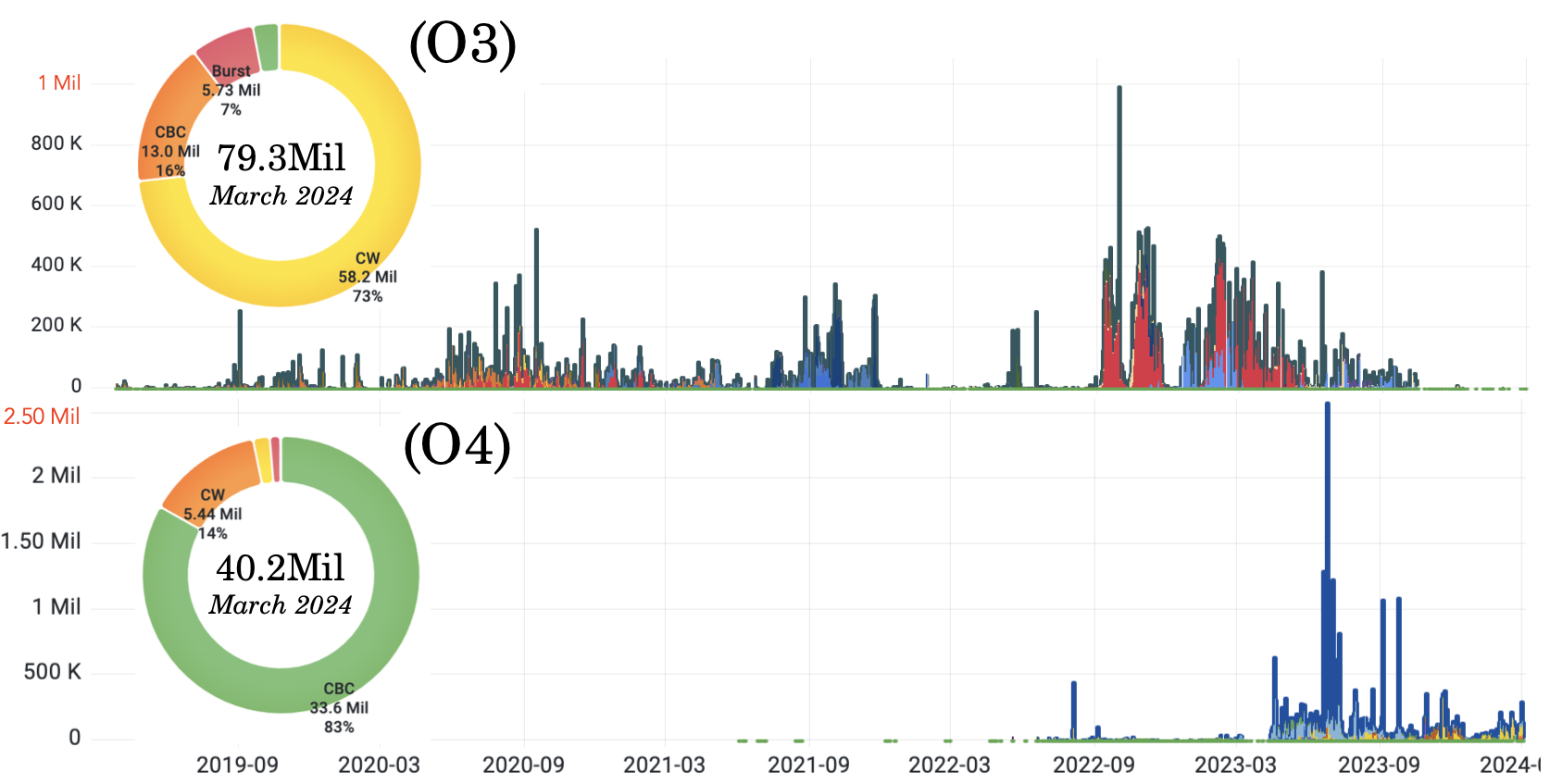}
    \caption{Comparison of O3 vs. O4 computing resources in March 2024. \cite{igwn_grafana}}
    \label{fig:o3-o4}
\end{figure}

Offline Analysis Infrastructure (or high-latency structure) is designed to provide a detailed and comprehensive examination of gravitational wave events that cannot be addressed in real time~\cite{bagnasco}. Similar to CERN infrastructures, such offline infrastructures support extensive computational and analytical tasks worldwide, providing valuable insights validating findings initially flagged by the Low Latency Alert Generation Infrastructure (LLAI) and more after continuous searches of refined signal candidates. Fig.\ref{fig:o3-o4} shows an order of magnitude in million CPU hours of computing resources used for O3 and the ongoing O4 run. Such a unit should be approximately compared with the CERN HEP benchmark unit as 1 CPU hour $\approx$ 10 MHS06~\cite{mhs06}. The comparison between O3 and O4 highlights the data's increasing volume and complexity of ground-based observational runs. Consequently, infrastructures are expanding to manage anticipated growing computing and storage resources in preparation for future observing runs.

Data analysis involved within the high latency infrastructure are typical signal extraction, noise reduction, detector characterization and detailed parameter estimations. Parameter estimation involves determining properties such as masses, distances, and spins of gravitational wave sources and is performed using sophisticated statistical methods and Bayesian inference techniques. Additionally, such infrastructure also serves as a platform for developing, testing new analytical methods, and experimenting with innovative techniques, improving the capabilities of the gravitational wave detection system and developing new modern and performant research pipelines.\\

A key component of this infrastructure is the use of Open Science Data Federation (OSDF)~\cite{osdf}, maintained by the Open Science Grid (OSG) consortium and currently deployed using CernVM-FS (CVMFS) protocol~\cite{cvmfs}. CVMFS is used for both software and data sharing in the current pattern. OSDF data \textit{origins} provide a distributed data storage and secured access, enabling the integration and management of large datasets across multiple institutions. Recently, the KAGRA collaboration enabled access to these resources by setting up an OSDF data origin server and cache on the KAGRA Main Data System. This federated approach facilitates seamless data sharing and collaboration, enhancing offline analysis capabilities by providing access to a broader range of data resources and computational power. 

\section{Data Analysis Framework Prototype}

The foundation of the gravitational wave analysis is carried out by the LALSuite, a C++ library initially developed by LIGO \cite{lalsuite}. It serves as the bedrock for gravitational wave (GW) low-latency analysis pipelines, signal processing, and parameter estimation. In elaborating on the integration of data analysis framework,  the ROOT Data Analysis Framework maintained by CERN \ cite {root, root-status} represents a natural and notorious candidate. A concrete example of its positive impact is demonstrated by the Omicron software~\cite{omicron}, which supports rapid computation of Q-Transform using ROOT data analysis framework in connection with LALSuite. \\

\begin{figure}[h]
    \centering
    \includegraphics[width=\textwidth]{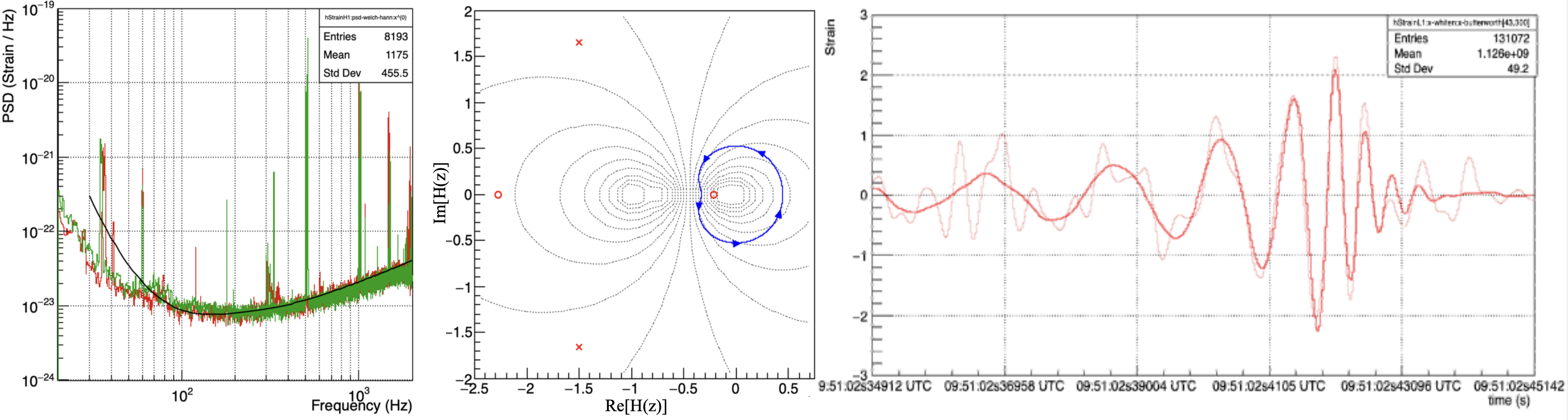}
    \caption{\textbf{(a)} Left: Power Spectrum Density calculation (GW150914~\cite{gw150914}); \textbf{(b)} Middle: example of digital filter $H(z)$; \textbf{(c)} Right: Matched filtering on c}
    \label{fig:signal-library}
\end{figure}

In this work, several prototyping libraries were developed, each serving distinct purposes. Among the most notable, there are a multi-format configuration file parser (incl. XML, YAML, JSON and INI format), allowing for flexible customization of software applications; a Kafka library interface, designed for efficient message passing; and real-time data handling and a few additional data formats parsing, such as HDF5 format~\cite{hdf5} and Frame Format \footnote{Proprietary data file format from GW community}~\cite{frameformat}. These are enhancing the compatibility of ROOT toward the GW community and bridging the gap between high-level programming interfaces and high-performance computational needs. In particular, a focus is now made on the introduction of advanced signal processing library including features such as illustrated in Fig.\ref{fig:signal-library}. These plots are known as typical examples extracted from gravitational wave analysis using Gravitational Wave Open Science Center (GWOSC) public data catalogs~\cite{gwtc-1, gwtc-2, gwtc-2-1}.

The integration of KFR library~\cite{kfrlib}, a modern C++ library actively maintained for around a decade, enables ROOT data analysis framework with two agnostic key features: rapid FFT calculations and advanced signal processing. KFR allows efficient handling of large signals through high-performance Fast Fourier Transform (FFT) computations in n-dimensions and also robust signal processing techniques, including windowing functions, Finite Impulse Response (FIR), and Infinite Impulse Response (IIR) filtering. 

\begin{figure}[h]
    \centering
    \includegraphics[width=1\textwidth]{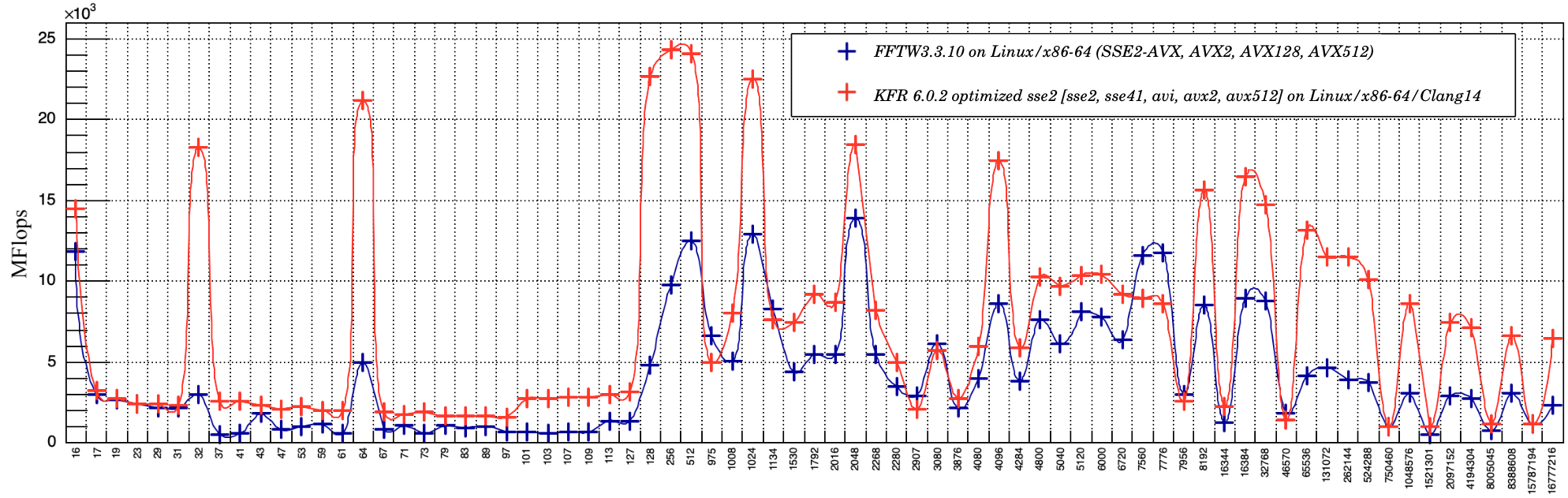}
    \caption{MFlops distributions (the higher, the better) vs FFT Size resulting from a benchmark test performed on a local cluster in Osaka, comparing complex forward transform (in-place) between FFTW 3.3.10 and KFR 6.0.2}
    \label{fig:kfrlib}
\end{figure}

Fig.\ref{fig:kfrlib} highlights the optimized performances of KFR over the well-known FFTW library~\cite{fftw}, commonly used in current IGWN and CERN implementations. KFR demonstrated enhanced performance in Fig.\ref{fig:kfrlib} based on FFT calculations at various FFT sizes and optimized for Intel AMD64 using Clang compiler. Application for benchmarking and reproducible results are available under \url{https://git.ligo.org/kagra/software/fft-benchmark} and original code was initiated by authors of the KFR library.

\section{Conclusions\label{sec:concl}}

The interplay between High-Energy Physics (HEP) and Gravitational Wave (GW) analysis has proven to be a powerful driver of scientific progress. The initial integration of HEP technologies in GW analysis and methodologies is part of a legacy from scientists originating from the HEP community in the 90s. Integrating modern technology might enhance gravitational wave observations' precision and scope and contribute to the HEP field. Continuous collaborations between HEP and GW communities would contribute to bridging gaps between particle physics and astrophysics. This ongoing interdisciplinary effort underscores the positive impact of integrating diverse complementary scientific approaches.\\

Following ACAT2024 conference, the aforementioned prototyping libraries were made available at \url{https://git.ligo.org/kagra/libraries-addons/root} and reorganized into a modular structure centred around ROOT+ library. This central library consolidates advanced signal processing by adding additional high-level helpers at no cost, implementing easy-to-use complex tensor calculations seamlessly on CPU and GPU, HTTP requests, and mainly relying on the LibTorch C++ API~\cite{libtorch}, also enabling for training and rapid inference of the most modern machine learning models by utilizing the Open Neural Network Exchange (ONNX) data format~\cite{onnx}. This integration of LibTorch within the prototyped ROOT+ library will provide a competitive option by seamlessly incorporating the most advanced developments in machine learning from open-source platforms while complementing the existing capabilities of the machine learning features already supported by TMVA SOFIE \cite{sofie} within ROOT Framework.

\section*{Acknowledgments}
This work was supported by MEXT, Grant-in-Aid for JSPS Fellows 22KF0329, JSPS Leading-edge Research Infrastructure Program, JSPS Grant-in-Aid for Specially Promoted Research 26000005, JSPS Grant-inAid for Scientific Research on Innovative Areas 2905: JP17H06358, JP17H06361 and JP17H06364, JSPS Core- to-Core Program A. Advanced Research Networks, JSPS Grantin-Aid for Scientific Research (S) 17H06133 and 20H05639 , JSPS Grant-in-Aid for Transformative Research Areas (A) 20A203: JP20H05854, the joint research program of the Institute for Cosmic Ray Research, University of Tokyo, National Research Foundation (NRF), Computing Infrastructure Project of Global Science experimental Data hub Center (GSDC) at KISTI, Korea Astronomy and Space Science Institute (KASI), and Ministry of Science and ICT (MSIT) in Korea, Academia Sinica (AS), AS Grid Center (ASGC) and the National Science and Technology Council (NSTC) in Taiwan under grants including the Rising Star Program and Science Vanguard Research Program, Advanced Technology Center (ATC) of NAOJ, and Mechanical Engineering Center of KEK.
This research was supported in part by the Japan Society for the Promotion of Science (JSPS) Grant-in-Aid for Scientific Research [Nos. 23H01176, 23K25872 and 23H04520 (H.\ Takahashi) and 22K03614 (K.\ Oohara)], Grant-in-Aid for JSPS Fellows [No. 22KF0329 (M.\ Meyer-Conde)] and Tokyo City University Prioritised Studies. This material is based upon work supported by NSF's LIGO Laboratory which is a major facility fully funded by the National Science Foundation. We also would like to express our deep gratitude to D. Cazarin, lead author of the KFR library, for extensive discussions and sharing the initial code of the benchmark test.

\printbibliography

\end{document}